# Evolution of seismic velocities in heavy oil sand reservoirs during thermal recovery process

# Évolution des vitesses sismiques dans les réservoirs de sables bitumineux au cours des procédés de récupération thermique

Nauroy, J.F, Doan, D.H., Guy, N., Baroni, A..
*IFP Energies nouvelles, 1-4 Av. du Bois Préau, 92852 Rueil-Malmaison Cedex, France*
Delage, P.
*Ecole des Ponts ParisTech, UMR Navier/CERMES, 6-8 Av. Blaise Pascal, F-77455 Marne la Vallée Cedex 2, France*
Mainguy, M.
*TOTAL Office EB-181 CSTJF, Av. Larribau 64018 Pau Cedex, France*

## ABSTRACT

In thermally enhanced recovery processes like cyclic steam stimulation (CSS) or steam assisted gravity drainage (SAGD), continuous steam injection entails changes in pore fluid, pore pressure and temperature in the rock reservoir, that are most often unconsolidated or weakly consolidated sandstones. This in turn increases or decreases the effective stresses and changes the elastic properties of the rocks. Thermally enhanced recovery processes give rise to complex couplings.

Numerical simulations have been carried out on a case study so as to provide an estimation of the evolution of pressure, temperature, pore fluid saturation, stress and strain in any zone located around the injector and producer wells.

The approach of Ciz and Shapiro (2007) - an extension of the poroelastic theory of Biot-Gassmann applied to rock filled elastic material - has been used to model the velocity dispersion in the oil sand mass under different conditions of temperature and stress. A good agreement has been found between these predictions and some laboratory velocity measurements carried out on samples of Canadian oil sand. Results appear to be useful to better interpret 4D seismic data in order to locate the steam chamber.

## RÉSUMÉ

Dans les procédés de récupération des huiles lourdes par méthodes thermiques, comme la stimulation cyclique par vapeur (CSS) ou le drainage par gravité assisté par vapeur (SAGD), l'injection de vapeur en continu entraine des changements de liquide de pores, de pression interstitielle et de température dans la roche réservoir, qui est constituée le plus souvent de sable non consolidés ou faiblement consolidés. Ces changements à leur tour augmentent ou diminuent les contraintes effectives et modifient les propriétés élastiques des roches. Les procédés de récupération par méthodes thermiques mettent en œuvre des couplages complexes.

Des simulations numériques ont été réalisées sur un cas d'étude afin de fournir une estimation de l'évolution des pressions, températures, saturation des fluides interstitiels, contraintes et déformations dans toute zone située autour des puits injecteur et producteur.

L'approche de Ciz et Shapiro (2007) - une extension de la théorie poroélastique de Biot - Gassmann appliquée aux roches remplies d'un matériau élastique - a été utilisée pour modéliser la dispersion de vitesse dans les sables bitumineux sous différentes conditions de température et de contrainte. Un bon accord a été trouvé entre les prédictions du modèle et certaines mesures de vitesse effectuées en laboratoire sur des échantillons de sables bitumineux Canadiens. Les résultats sont utiles pour mieux interpréter les données sismiques 4D afin de localiser la chambre à vapeur.

## INTRODUCTION

[1] Huge quantities of heavy oils (heavy oil, extra heavy oil and bitumen) are mainly trapped in unconsolidated sand and sandstone reservoirs in Western Canada and Eastern Venezuela basins. In thermally enhanced recovery processes like cyclic steam stimulation (CSS) or steam assisted gravity drainage (SAGD), the injection of steam in oil sand deposits produces changes in temperature, pressures and pore fluid content. These changes obviously affect the elastic and seismic properties of the soil layers. Time-lapse 3D seismic (4D seismic) and other seismic technologies can be used to monitor the impact of these changes in unconsolidated sands or weakly consolidated sand reservoirs (Zhang *et al*., 2005; Schmitt, 2005; Nakayama *et al.*, 2008; Tanaka *et al.*, 2009). By comparing maps of seismic attributes (velocity, amplitude, attenuation, etc) at different times, the continuous spatial distribution of the heated reservoir zones can be approximately located. So, the feasibility of seismic monitoring methods is based on the visible changes of seismic attributes during thermal process that could be understood from a combination of rock physics modelling and direct laboratory measurements.

[2] Over the past three decades, several laboratory measurements of seismic properties (wave velocity and attenuation) of heavy oil sands have been published (Mraz *et al*., 1982; Nur *et al.*, 1984; Tosaya *et al*., 1987; Wang and Nur, 1988, 1990; Wang *et al*., 1990; Eastwood, 1993; Kato *et al*., 2008). However, there is still a lack of clear understanding due on the one hand to well-known difficulties in obtaining representative samples of oil sands and on the other hand to the limits of testing systems generally designed for hard rock, but not for unconsolidated oil sands. In addition, the application to field problems of the results obtained in the laboratory with ultrasonic frequency is not obvious because of the wave dispersion phenomenon occurring in heavy oil sands.

[3] Several attempts to model the properties of rock saturated with heavy oil have been made (Das and Batzle, 2008; Ciz *et al*., 2009). The difficulties are largely related to the viscoelastic behaviour of heavy oils. They behave like a quasi-solid at high frequencies and/or low temperature and like a viscous fluid at low frequencies and/or high temperature (Batzle *et al*., 2006b; Hinkle, 2008). Heavy oils have a non-negligible shear modulus that depends on both frequency and temperature and that allows propagation of shear waves. So, the physical characterization of rock saturated with heavy oil is different from that of rock saturated with conventional fluids with no shear modulus. It makes the standard poroelastic Biot-Gassmann approach (Biot, 1941; Gassmann, 1951) inapplicable (Ciz and Shapiro, 2007; Gurevich *et al*., 2008; Ciz *et al*., 2009). The generalization of the Biot-Gassmann equations (Ciz and Shapiro, 2007), originally made for porous rock saturated with an elastic material, can be considered as the first step to model the seismic responses of heavy oil-saturated rocks. The feasibility of this approach in predicting the velocities of heavy oil sand has been considered in this paper.

# 1. EVOLUTION OF PRESSURE, TEMPERATURE AND PORE FLUID SATURATION DURING SAGD

[4] The coupled thermo-hydro-mechanical modelling of SAGD has been conducted at IFPEN (Lerat *et al*., 2010; Zandi, 2011), showing how the different fields (stress, pressure, temperature and steam) develop with respect to time. The result presented in this section has been obtained through a coupled simulation made by using the geometry and the parameters already presented by Zandi (2011). This simulation is based on an iterative coupling between a reservoir simulator (PumaFlow$^{TM}$) and a geomechanical simulator (ABAQUS$^{TM}$) in order to take into account the geomechanical effects on the porous volume and the permeability changes of the reservoir. In this simulation, a 20 meters thick (in the *Z* vertical direction) reservoir at a depth of 250 meters is modelled. The domain considered for the reservoir simulator is rectangular with a width of 147 meters (in the *X* direction) and a length of 500 meters along the *Y*-axis of the well pair. The mesh used by the reservoir simulator in the *X-Z* plane is shown in figure 1a. It contains only one grid block in the *Y* direction. The pair of wells is located at the centre of the domain and the injection well is 3.5 meters upper than the production well (see figure 1a where points *I* and *P* represent the position of the injection and production wells, respectively). The domain considered in the geomechanical model is built from the surface to a depth of 280 meters, including an underburden layer 10 meters thick. It should be noted that, in the geomechanical model, the displacements along the *Y* axis are blocked, corresponding to a plane strain condition. For simplicity, an elastic and isotropic behaviour is considered for the rock. More information about the parameters considered has been given by Zandi (2011).

[5] The simulations are performed over 1500 days. The first 120 days correspond to the pre-heating of the regions surrounding both wells by circulating steam in order to reduce the oil viscosity and create a hydraulic link between them. Once pre-heating is finished, steam injection starts with a temperature of about 260°C. The steam injection and production rates are automatically controlled with a maximal pressure sets to 5 MPa in the injection well and a minimum pressure of 0.5 MPa for the production well. Rates are automatically controlled based on the analysis of the temperature of each well. The rates are adjusted in order to keep the production well temperature between 20°C and 35°C lower than the injection well temperature.

[6] In the model associated with the reservoir simulator, fluids cannot flow through the boundaries, but heat losses by conduction through the upper and lower boundaries are taken into account by a simplified one-dimensional modelling of the overburden and underburden that is oriented in the vertical direction. In the geomechanical model, the horizontal displacements of lateral boundaries, as well as all the displacements of the lower boundary, are blocked.

[7] The iterative coupling between the reservoir simulator and the geomechanical simulator is performed for periods of 5 days. It means that the reservoir permeability and porous volume are corrected 300 times during the simulations.

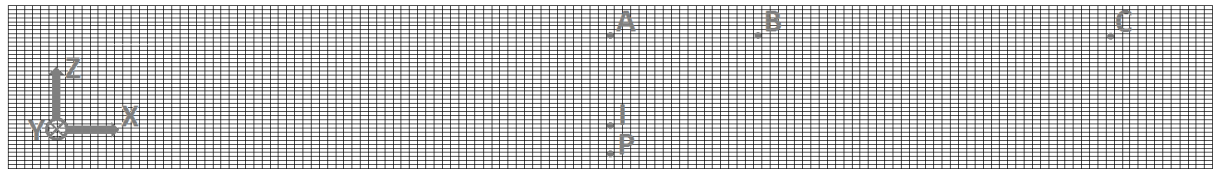

a) Geometry modelling and location of the grid cells for the reservoir simulator and definition of grid cells *A*, *B* and *C*

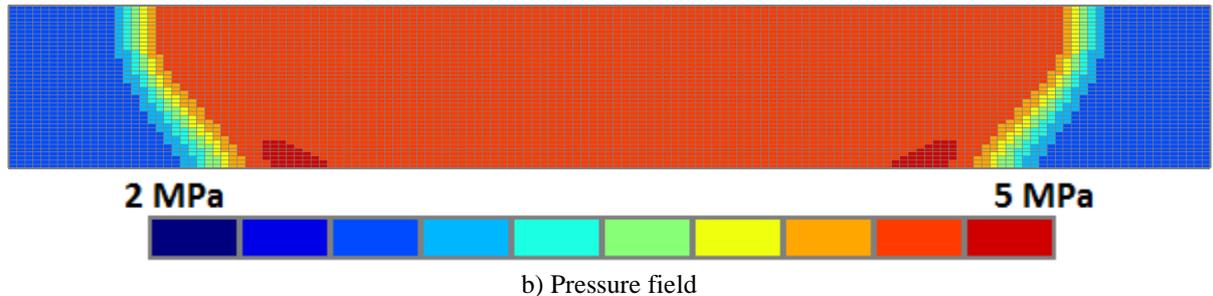

b) Pressure field

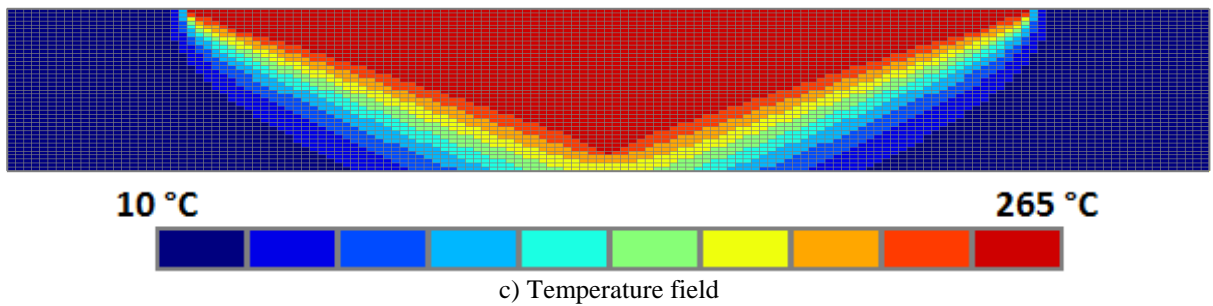

c) Temperature field

Figure 1. Definition of reservoir simulator spatial discretization (a) and pressure (a) and temperature (c) fields at the end of simulation (1500 days) in oil sand reservoir (at 250 m depth).

[8] The evolution of the main variables has been studied in three different grid cells which represent three different zones in the reservoir (Figures 2, 3 and 4). These cells are located in the same *X-Z* plane and at the same elevation in the *Z* direction. Figure 1a shows the location of these grid cells (*A*, *B* and *C*). Grid cell *A* is placed just in the middle of the reservoir and 5.5 meters above the injection well. Grid cells *B* and *C* are chosen farther from the wells, with the distance of 18.5 and 60 meters from the grid cell *A*, respectively.

[9] Inside the reservoir, we can clearly see the arrival of different fronts (very visible in *B*). The pore pressure front is firstly followed by the temperature front and, at the end, the change of saturation.

[10] The invasion of steam (consisting of at least 70 % vapour water with the remainder in the liquid phase) occurs when temperature exceeds 70°C, firstly by replacing oil with water, then with steam vapour when temperature reaches 260°C. Accordingly, at grid cell *B*, located within the steam chamber the pore space is filled after 1500 days of steam injection with a mixture of 72 % steam vapour, 20 % water and 8 % oil. Temperature will be about 260°C.

[11] The formation of the steam chamber leads to effective and total stress changes and strain of the reservoir rock that cause permeability modifications (Touhidi-Bagigni, 1998; Dusseault, 2008). Figures 2, 3 and 4 illustrate the changes in vertical ($\sigma_{zz}$) and horizontal ($\sigma_{xx}$ and $\sigma_{yy}$) stresses for grid cells *A*, *B* and *C* respectively (compression stress taken positive). The stress state at grid cell A is significantly changed when the steam injection starts (after 120 days of pre-heating). The steam injection leads to an increase in temperature and pore

pressure around the wells. The vertical total stress begins to increase at grid cell *A* whereas it begins to decrease for grid cells *B* and *C*. This phenomenon can be related to a reversed arching effect. Indeed, the rock expands around the wells because of temperature and pressure changes, leading to a vertical total stress increase around the wells and at grid cell *A*. When the pressure front reaches grid cell *B* (after about 400 days as shown in figure 3), the total vertical stress in *B* starts to increase. This increase is afterwards enforced with the arrival of the temperature front after about 600 days. At the end of the simulation, the pressure and temperature fronts have not clearly reached grid cell *C* and this is the reason why the total vertical stress decreases only at this grid cell.

[12] The horizontal total stress $\sigma_{xx}$ in the reservoir clearly increases during the steam injection, in relation with the lateral boundary conditions adopted in the geomechanical model. The total horizontal stress is changed accordingly except at grid cell *A* (*i.e.* around the wells) at the beginning of steam injection (figure 2) when the steam chamber lateral extension is small compared to the reservoir width.

[13] The evolution of the other total horizontal stress $\sigma_{xy}$ is strongly related to the plane strain hypothesis adopted in the geomechanical model; it increases at every stages of production in all grid cells (*A*, *B* and *C*).

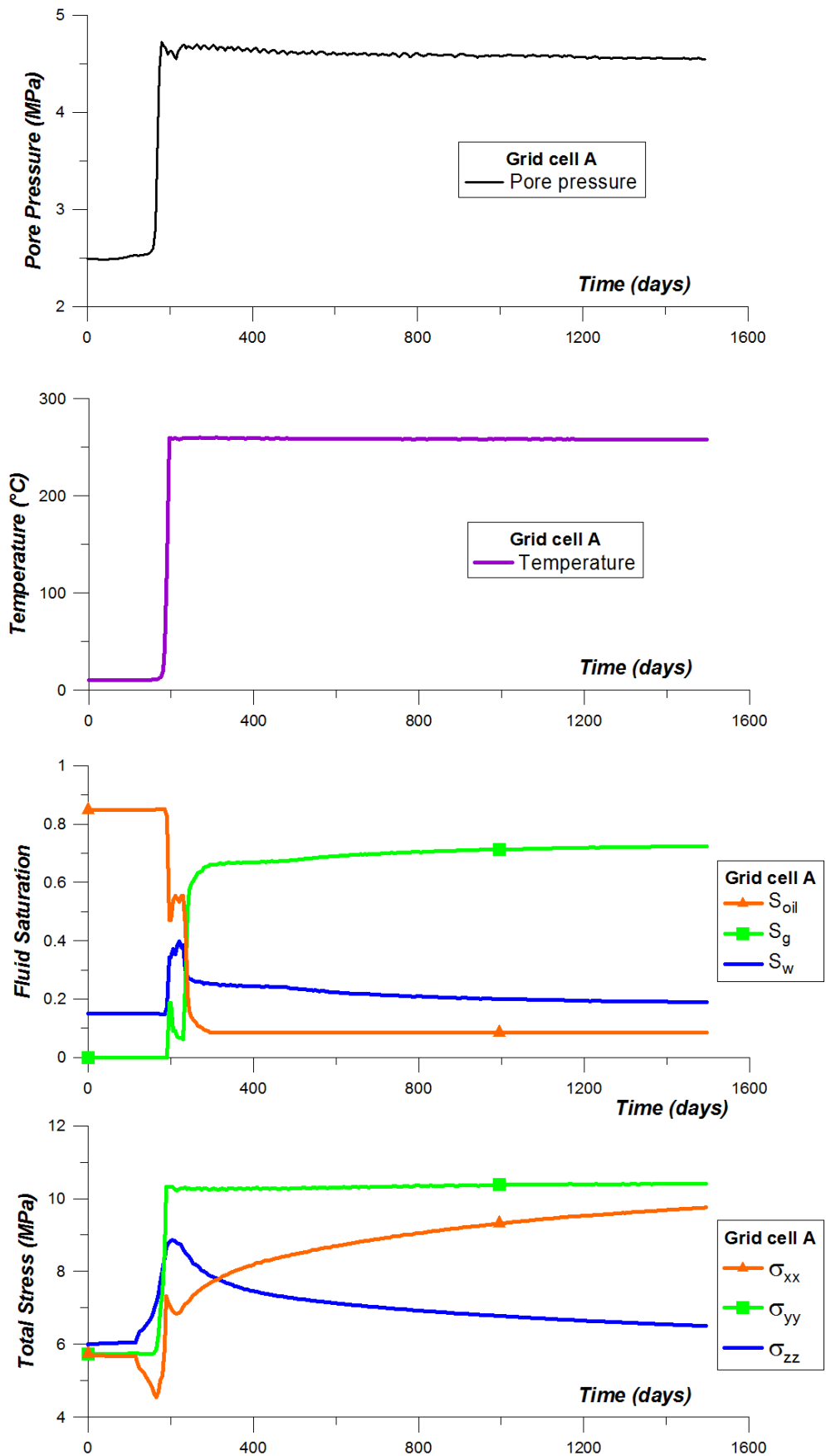

Figure 2. Evolution of pore pressure, temperature, pore fluid saturation and total stresses during simulation at grid cell *A*.

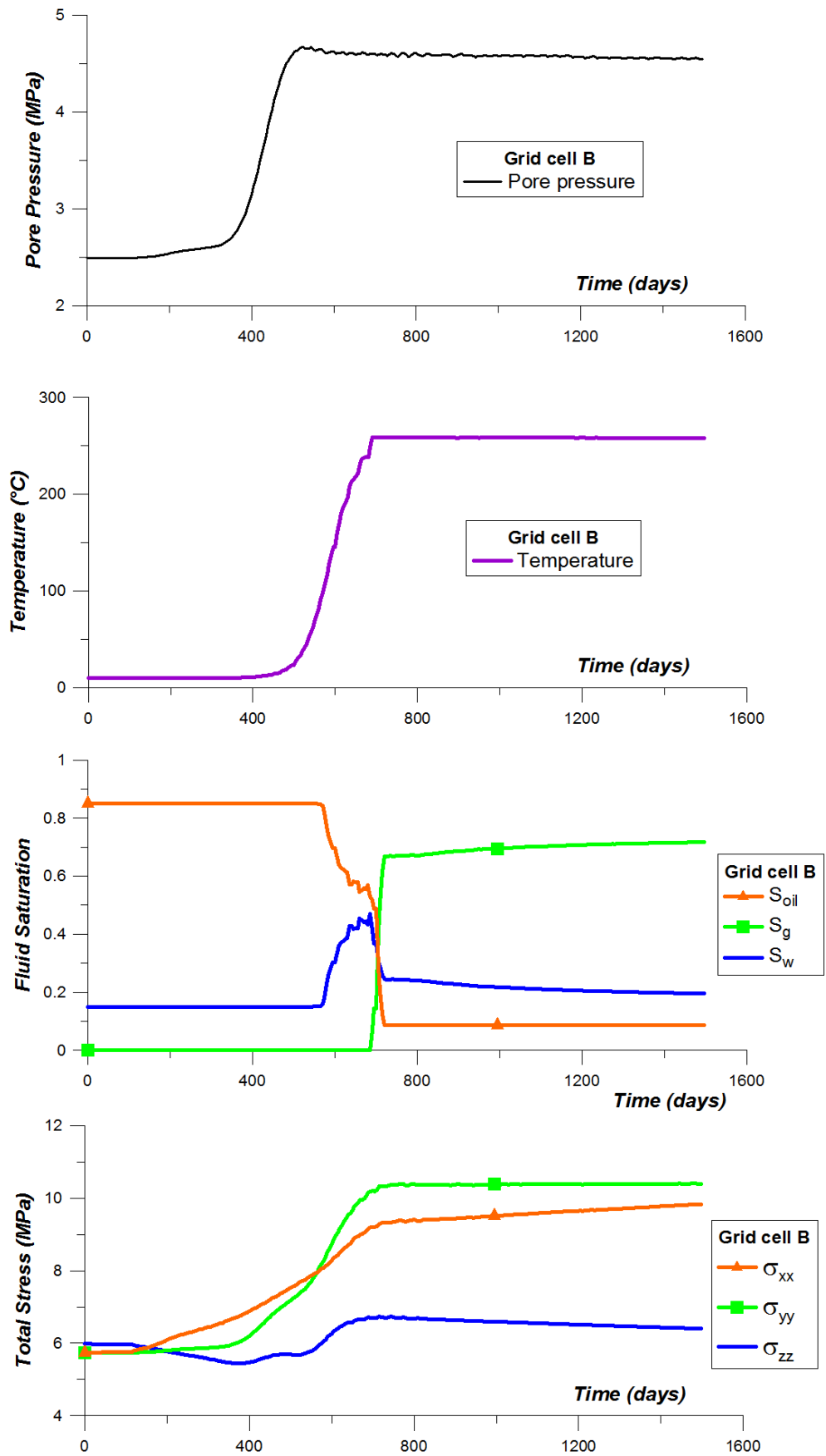

Figure 3. Evolution of pore pressure, temperature, pore fluid saturation and total stresses during simulation at grid cell *B*.

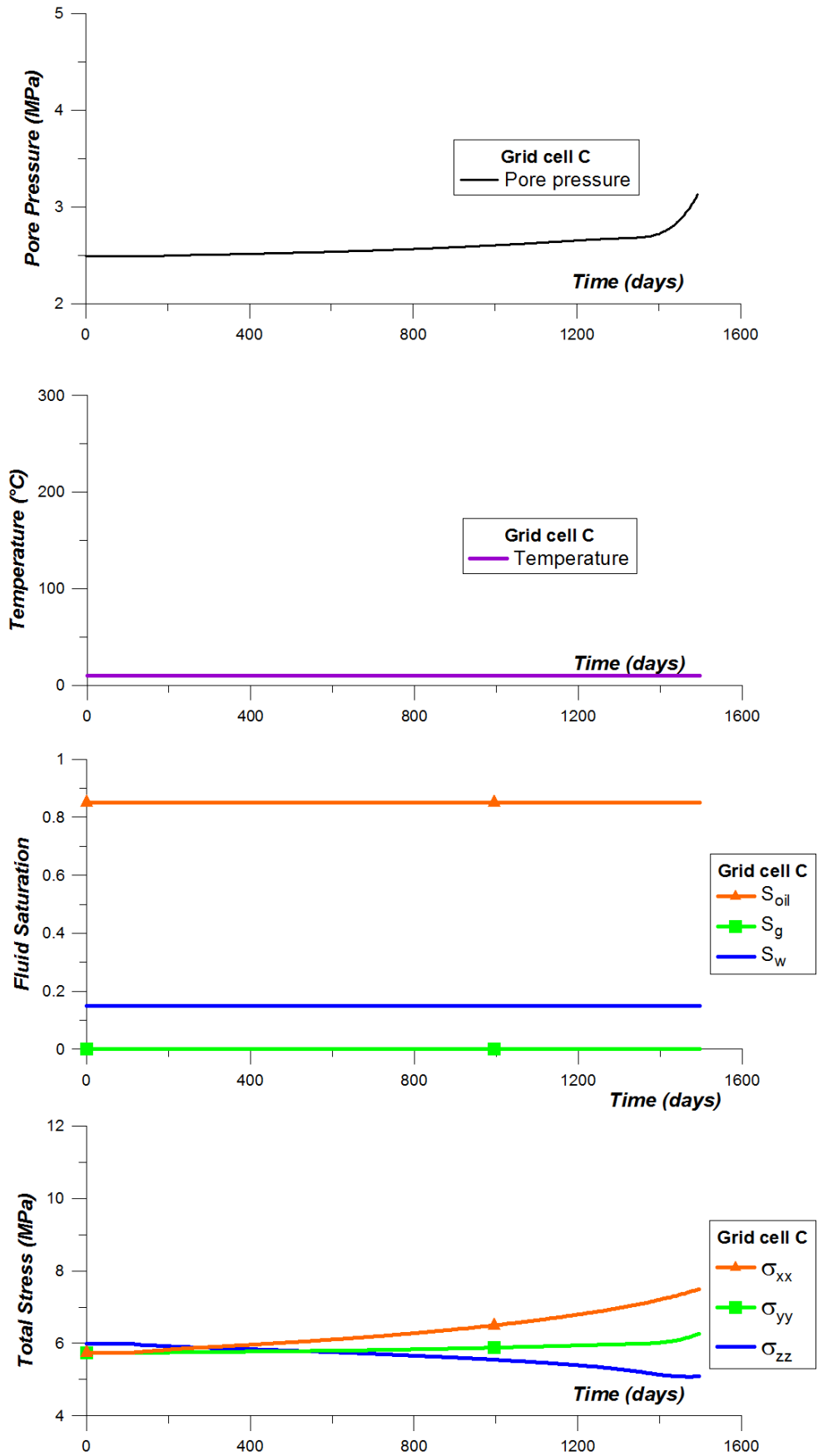

Figure 4. Evolution of pore pressure, temperature, pore fluid saturation and total stresses during simulation at grid cell *C*.

[14] The geomechanical computation results are also presented in a *p'-q* diagram (mean effective stress – deviatoric stress) in figure 5 for grid cells *A*, *B* and *C*. Considering that the reservoir is confined, two schematic tendencies can be pointed out. For a rock element, firstly, a temperature increase will preferentially lead to an increase of *p'* and to an increase of *q* from an isotropic initial stress state. Secondly, a pore pressure increase will preferentially lead to a decrease of *p'* and an increase of *q* from an isotropic initial stress state.

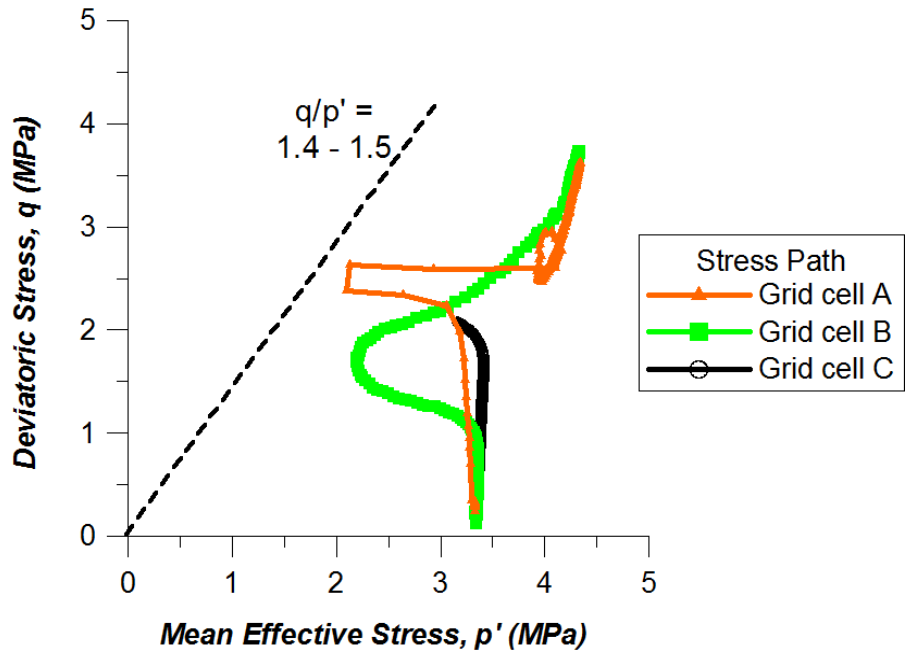

Figure 5. Stress paths in grid cells *A*, *B* and *C* after 1500 days of simulation.

[15] As shown in figure 5, the stress path of each grid cell is different, especially far from the steam chamber (grid cell *C*). The same tendency can be observed for grid cells *A* and *B*: starting from an isotropic stress state, *p'* decreases and *q* increases when the pore pressure front arrives. *p'* and *q* afterwards increases when the temperature front arrives. For grid cell *C*, after 1500 days, the pressure front just arrived and but not the temperature front.

[16] A typical plastic limit characterised by $q/p' = 1.4 – 1.5$ (see Doan, 2011) is also drawn in figure 5. This plastic limit is close from being reached and it could probably be reached starting from another initial stress state, for example $p' = 2$ MPa and $q = 1$ MPa. The change in both *p'* and *q* appear significant, compared to the plastic limit. It should be pointed out that the stress state of grid cells *A* and *B* is closer to the plastic limit once the pore pressure front passed and while the temperature front is arriving.

## 2. VISCOELASTIC MODEL FOR CANADIAN OIL SAND

[17] The elastic properties of the oil sands depend on the properties of the sand matrix, on the nature of the pore fluid (bitumen, water or steam) and of course on in situ stresses and pore pressure. These properties also depend on the frequency of the elastic waves since oil sands are viscoelastic materials.

[18] The approach of Ciz and Shapiro (2007), originally developed for porous rocks saturated with an elastic material, has been used to compute the effective elastic properties of porous rocks filled with various fluids including heavy oil, a viscoelastic material with a non negligible shear modulus. This approach reduces to the classical Biot-Gassmann one if the

pore fluid is non-viscous in the case of a rock skeleton made up of a single homogeneous mineral.

[19] According to Ciz and Shapiro, the two following equations describe the effective moduli of a sand saturated with a viscous fluid:

$$K_{sat}^{-1} = K_{dr}^{-1} - \frac{(K_{dr}^{-1} - K_s^{-1})^2}{\phi(K_f^{-1} - K_s^{-1}) + (K_{dr}^{-1} - K_s^{-1})} \quad \text{(Eq. 1)}$$

$$G_{sat}^{-1} = G_{dr}^{-1} - \frac{(G_{dr}^{-1} - G_s^{-1})^2}{\phi(G_f^{-1} - G_s^{-1}) + (G_{dr}^{-1} - G_s^{-1})} \quad \text{(Eq. 2)}$$

where $\phi$ is an uniformly distributed porosity.

$K_{dr}, G_{dr}$ are the drained bulk and shear moduli ($K_{dr} = K_{dry}$, $G_{dr} = G_{dry}$) of the clean sand (with no oil), respectively.

$K_{sat}, G_{sat}$ are the effective bulk and shear modulus of the undrained saturated system.

$K_f, G_f$ are the bulk and shear moduli of saturating fluid.

$K_s, G_s$ are the moduli of the solid phase.

[20] The sketch of the modelling process is presented in the figure 6.

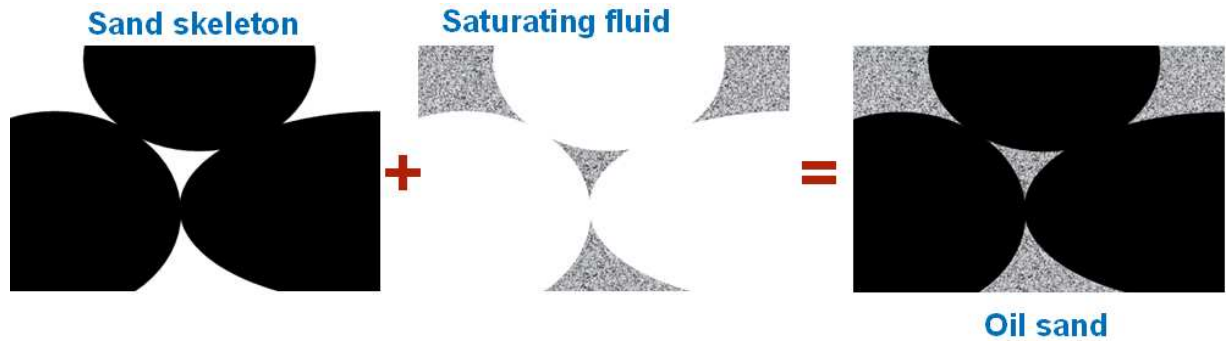

Figure 6. Sketch of modelling process for Canadian oil sand (Ciz and Shapiro approach).

## 2.1. Moduli of the sand skeleton

[21] The drained moduli of the unconsolidated or weakly cemented reservoir formation have been widely investigated in the literature. They are independent of the frequency (Spencer *et al.*, 1994; Batzle *et al.*, 2006a), of temperature (Wang and Nur, 1990; Eastwood, 1993; Batzle *et al.*, 2006a; Doan, 2011) and highly stress dependent.

[22] Here, the elastic moduli of the sand skeleton were computed by using a Hertz-Mindlin contact model widely applied in dry granular media. The effective elastic moduli of a dry, dense and random pack of identical elastic spheres is given by (Mavko *et al.*, 1998):

$$K_{dr} = \frac{C(1-\phi)}{12\pi R_g} S_n \quad \text{(Eq. 3)}$$

$$G_{dr} = \frac{C(1-\phi)}{20\pi R_g}(S_n + \frac{3}{2}S_t) \quad \text{(Eq. 4)}$$

where $C$ is the average number of contacts per grain (coordination number), $R_g$ is the mean grain radius and $\phi$ is the porosity.

[23] Number $C$ was estimated by using the following empirical function $C = 20 - 34\phi + 14\phi^2$ (Avseth *et al.*, 2010).

[24] $S_n$, $S_t$ are the normal and tangential stiffness, respectively. The normal stiffness $S_n$ between two identical spheres in the pack under a confining pressure $P$ (Mavko *et al.*, 1998; Bachrach *et al.*, 2000) is given by:

$$S_n = \frac{4aG_s}{1-\upsilon_s} \quad \text{(Eq. 5)}$$

[25] The tangential stiffness $S_t$ of two identical spheres under a confining pressure $P$ should be zero if a perfect slip occurs between the spheres. If there is no slip at the contact surface (finite friction), this stiffness is given as (Walton, 1987; Mavko *et al.*, 1998):

$$S_t = \frac{8aG_s}{2-\upsilon_s} \quad \text{(Eq. 6)}$$

where $G_s$ and $\upsilon_s$ are the shear modulus and the Poisson's ratio of the solid grains, respectively, $a$ the radius of the contact area,

$$a = \sqrt[3]{\frac{3FR_c(1-\upsilon_s)}{8G_s}} \quad \text{(Eq. 7)}$$

where $R_c$ is related to the local radii of curvature of two grains $R_1$ and $R_2$ (Bachrach *et al.*, 2000); and $F$ is a confining force acting between two particles,

$$F = \frac{4\pi R_g^2 P}{C(1-\phi)} \quad \text{(Eq. 8)}$$

[26] Some particular patterns could be noted in our application. Firstly, the introduction of parameter $R_c$ (see Equation 7) allows to take into account the grain angularity, a feature clearly observed from microtomography images of Canadian oil sands (Doan, 2011). As quoted by Zimmer *et al.* (2007), the non-spherical shape of sand grains is one of main reason of the difference between empirical and theoretical values. We have assumed $R_c = 0.1 \times R_g$. Secondly, as pointed out by Zimmer *et al.* (2007), the friction of two grains in contact is in reality neither infinite nor zero, and the tangential stiffness depends on the boundary conditions and of the loading path (Bachrach and Avseth, 2008). Consequently, we considered here a random case where there are 50% slipping grains and 50% no slipping grains. The Hashin-Strikman average bound has been afterward used to compute the tangential stiffness $S_t$ of the mixture.

## 2.2. Moduli of the saturating fluid

[27] We considered a three-component fluid phase, heavy oil (with saturation $S_{oil}$), water (with saturation $S_w$) and gas ($S_g$). Since there is no gas phase in oil sands under in-situ conditions (Dusseault, 1980), the introduction of gas occurs as a result of steam injection.

[28] Heavy oils exhibit a viscoelastic behaviour with complex moduli that depend both on frequency and temperature (Batzle *et al.*, 2006b; Hinkle, 2008). The Cole-Cole model (see Gurevich *et al.*, 2008) was adopted for the complex shear modulus $G_{oil}$:

$$G_{oil} = G_0 + \frac{G_\infty - G_0}{\frac{1}{(-i\varpi\tau)^\beta} + 1} \qquad \text{(Eq. 9)}$$

where $G_0$ and $G_\infty$ are the shear moduli at zero frequency (very slow deformation) and infinity frequencies (quick deformation), respectively. Exponent $\beta < 1$ is an adjustable parameter. $\tau = \eta / G_\infty$ is the relaxation time and $\eta$ is the viscosity.

[29] The empirical relationship between viscosity and temperature suggested by Gurevich *et al.* (2008) was used:

$$\ln(\frac{\tau}{\tau_\infty}) = A . \exp(-\frac{T}{T_0}) \qquad \text{(Eq. 10)}$$

where $\tau_\infty = \eta_\infty G_\infty^{-1}$,
$T$ is temperature in Celsius degree,
and $A$, $T_0$ are adjustable parameters.

[30] By default in the model, the viscosity value does not reduce below $\eta_\infty$. Examination of the changes of heavy oil viscosity curve with respect to temperature provides a minimum viscosity value $\eta_\infty = 10^{-3} Pa.s$.

[31] The complex bulk modulus of heavy oil is computed by the following relation (Ciz *et al.*, 2009):

$$K_{oil} = K_{ref} + \frac{5}{3} \times G_{oil} \qquad \text{(Eq. 11)}$$

where $K_{ref}$ is the bulk modulus of the free non-viscous fluid ($K_{ref}$ = 2.22 GPa).

[32] Ignoring the wetting effect of the mixed fluid phase (oil, water and gas), the overall fluid bulk modulus has been calculated using Ruess average (Mavko *et al.*, 1998):

$$\frac{1}{K_f} = \frac{S_{oil}}{K_{oil}} + \frac{S_w}{K_w} + \frac{S_g}{K_g} \qquad \text{(Eq. 12)}$$

where $K_{oil}$, $K_w$ and $K_g$ represent the bulk moduli of the separate phases of oil, water (*w*) and gas (*g*) with respective degree of saturations $S_{oil}$, $S_w$ and $S_g$.

[33] The overall fluid shear modulus has been calculated using:
$$G_f = G_{oil} \text{ if } S_w + S_g \neq 1 \qquad \text{(Eq. 13)}$$
$$G_f = 0 \text{ if } S_w + S_g = 1 \qquad \text{(Eq. 14)}$$

[34] The overall fluid density has been calculated using:
$$\rho_f = S_{oil}\rho_{oil} + S_w\rho_w + S_g\rho_g \qquad \text{(Eq. 15)}$$
where $\rho_{oil}$, $\rho_w$ and $\rho_g$ represent the bulk moduli of oil, water (*w*) and gas (*g*).

## 2.3. Calibration of the model on laboratory experiments

[35] Laboratory measurements of *P*- and *S*-waves velocities and attenuations have been performed on natural oil sand samples coming from fluvial-estuarine McMurray sand cores extracted at 75 m below surface from a shallow Athabasca deposit. The in situ porosity of the oil sand specimen has been estimated in the range 31 - 35 % from petrophysical log. Oil sands

are mostly made of quartz, with grain moduli of $K_s = 38$ GPa, $G_s = 44$ GPa and a grain density of 2.65 Mg/m$^3$.

[36] Velocity measurements have been performed in a pressure cell at 0.5 MHz and in a triaxial cell at 1 MHz under different temperatures (up to 160°C) and isotropic effective stress. Details can be found in Doan *et al.* (2010) and Doan (2011). In situ velocities obtained by sonic log measurement (frequency 10 kHz) were also available.

[37] Some main experimental results can be briefly enumerated as follow. *P*- and *S*-wave velocities showed a continuous decrease with increasing temperature. The decrease appeared more pronounced at temperature lower than 60°C. Moreover, velocities gradually increased with increasing confining pressure, as commonly observed in rocks. Sample saturation with a viscous fluid increased both the *P*- and *S*-wave velocities. This can be interpreted not only as a direct consequence of the viscosity effect but also as a result of the viscous contribution to the bulk moduli under saturated conditions.

[38] In addition, our measured high attenuation values on oil sand suggested significant energy losses in the heavy oil sand during the propagation of an acoustic wave. *P*- and *S*-wave attenuations showed a peak as a function of temperature or of the viscosity of the saturating fluid.

[39] These changes in velocity and attenuation reflect the fact that the oil sand is an attenuating and as a consequence, dispersive medium. Therefore the wave velocities and attenuations depend on the frequency. This is extremely important because laboratory measurements are made at 0.5 or 1 MHz whereas the frequency of the seismic waves used in-situ are much smaller, around 100 Hz.

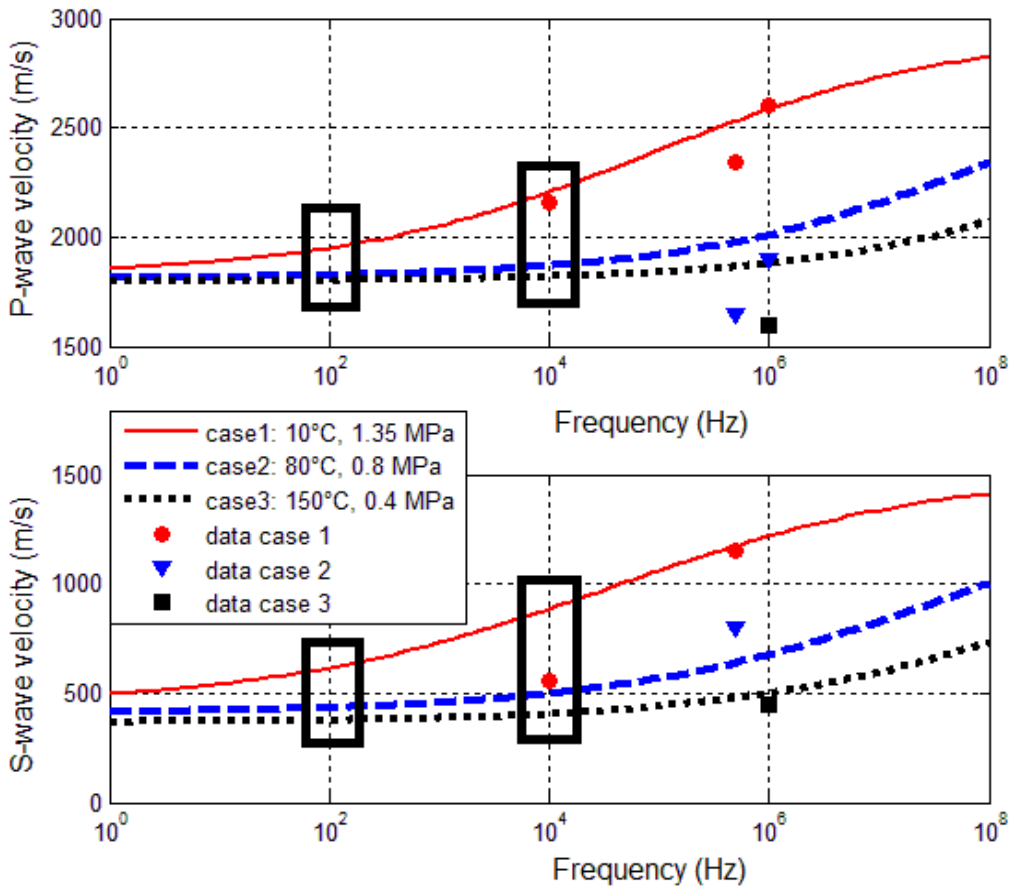

Figure 7. Modelling of velocity dispersion in oil sand. The points around 10$^4$ Hz represent in situ velocities and the points around 10$^6$ Hz represent our measured data.

[40] The model developed here was used to compute the velocity dispersion in three theoretical cases (figure 7). Case 1 corresponds to the initial reservoir conditions (i.e. shallow Athabasca reservoirs) in which the temperature is about 10°C and the in situ stress about 1.35 MPa (corresponding to a depth of about 75 m). Case 2 (80°C and 0.8 MPa) and case 3 (150°C and 0.4 MPa) correspond to the occurrence of steam injection with increases in temperature and pore pressure. A comparison between the predicted and experimental results (i.e. both field velocities and laboratory-measured velocities) (figure 7) shows a good match, especially for the *P*-wave velocity. It should be noted that the *S*-wave velocity at 10 kHz appeared significantly low in comparison with modelling results, but the very high *Vp/Vs* (e.g. around 3.9) casts doubt on the real value of *Vs*.

[41] The modelling results showed some other interesting patterns:
o  The velocity dispersion (i.e. the frequency dependency with respect to velocity) is highlighted at all the conditions of temperature and pressure considered;
o  The velocity dispersion curves shift to higher frequencies as temperature increases.

[42] For a fixed frequency (figure 7), the predicted velocities considerably change when passing from case 1 (10°C, 1.35 MPa) to case 3 (150°C, 0.4 MPa). At a seismic frequency of about 100 Hz, a decrease of 10% for *Vp* and of 30% for *Vs* could be expected. At a sonic frequency of around $10^4$ Hz, *P*-wave velocities drop by about 18% and *S*-wave velocities drop up to 44%. Further measurements are needed to check the computed *S*-wave velocities.

## 3. APPLICATION FOR TIME-LAPSE MONITORING

[43] According to the previous experimental and modelling results, the probable evolution of *P*- and *S*-wave velocities in an oil sand mass (Marc: deposit ou reservoir?) at seismic frequency bandwidth (cf. 100 Hz) at grid cell *B* (located in figure 1) is illustrated in figure 8 with respect to the arrival of the different fronts associated to steam injection. The seismic velocities have been calculated by Ciz and Shapiro approach previously described, by using input values of pore pressure, temperature and fluid saturation deduced from SAGD coupled thermo-hydro-mechanical modelling.

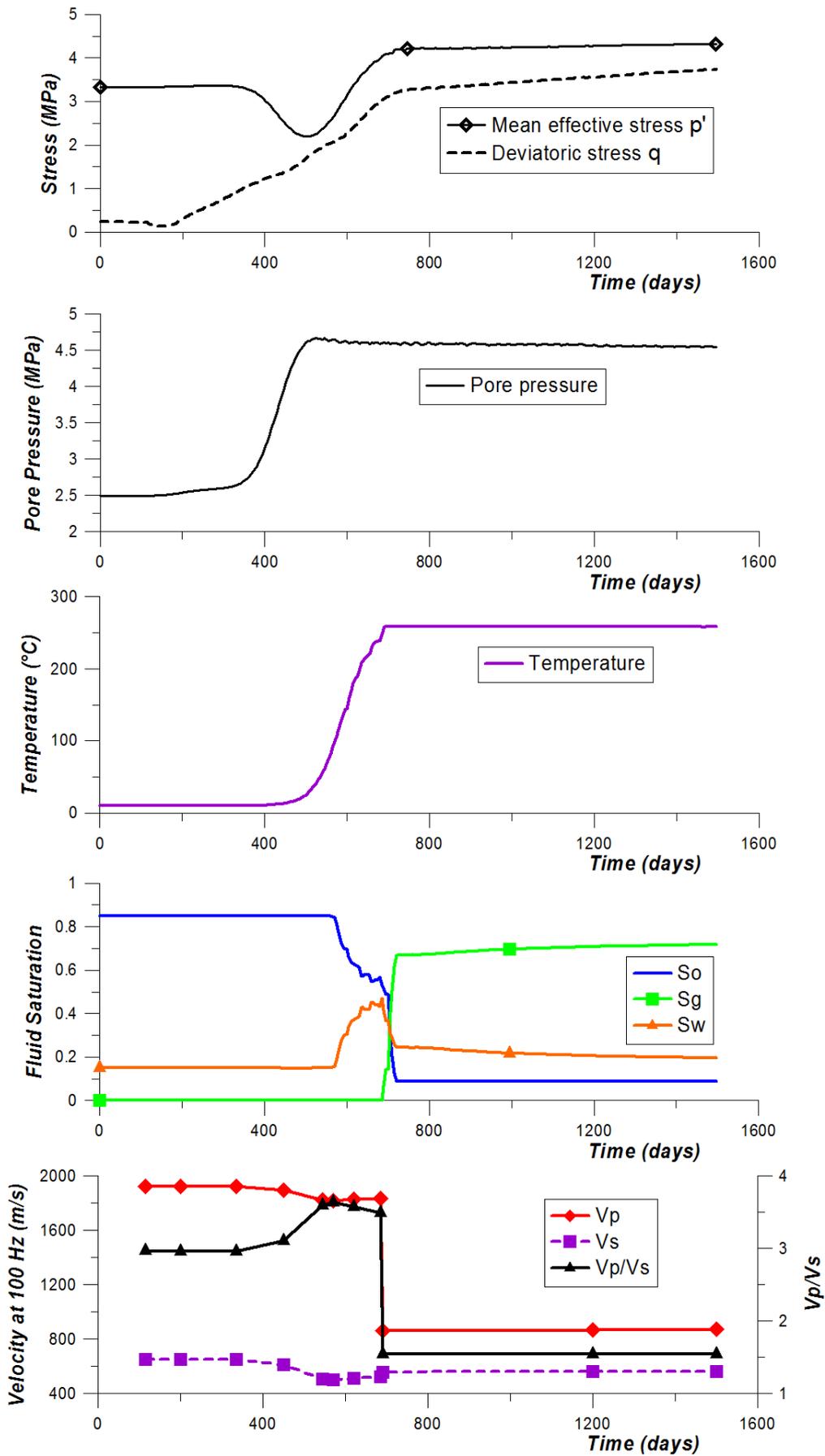

Figure 8. Evolution of the different fields at a typical grid cell (*B*) in an oil sand reservoir (250 m depth)

[44] The arrival of the first front called "stress" is related to a structural effect. As the mean total stress remains constant, it was assumed that the velocities do not change because we only take into account velocity dependence on mean stress in our model, and as a consequence, the velocity is assumed isotropic. Further work should be done to include the effect of deviatoric stress changes on velocities. In this case, the velocities should probably increase in the direction of the maximum total stress.

[45] When the pore pressure front arrives, velocities *Vp* and *Vs* gently decrease, as the mean effective stress decreases.

[46] As the temperature front arrives, the velocities decrease of 10 % for the compressional waves and of 30 % for shear waves. The decrease appears more pronounced below 60°C, levelling off beyond this temperature level. These modelling results are consistent with experimental studies reporting a noticeable change in both velocities at low temperature (cf. below 60°C). Indeed, because the amount of decrease of *S*-wave velocity is relatively larger than that of *P*-wave velocity, the *Vp/Vs* ratio significantly increases.

[47] The substitution of heavy oil and water by steam (at around 260°C) drastically reduces the compressional velocities *Vp* with little effect on the shear velocities *Vs*. The introduction of even a small percentage of gas bubbles into the fluid will have a significant effect on the bulk modulus of the mixture but very little effect on its density (see Equation 12 and 15). The sudden appearance of gas has little effect on *Vs* due to decreasing density. The *Vp/Vs* ratio thus drops significantly.

[48] In summary, it is observed that velocities decrease with the steam arrival. These successive changes could be identified by 4D seismic. The sudden appearance of gas is the more influent factor in changing the velocities. It should be pointed out that the evolution of *S*-wave velocities with respect to temperature are relatively more pronounced than that of *P*-wave velocities, suggesting to use the *S*-wave velocities as a indicator of temperature. It should also be noted that at seismic frequencies (up to 100 Hz), most changes in velocities occur at low temperature (below 60°C) with only small changes in velocities above this temperature. In other words, one could practically locate the heated zones in the reservoirs but one probably could not deduce the exact temperature of these zones. One would also expect a possible initial decreasing attenuation at seismic frequencies (especially in deep reservoirs) followed by an increasing attenuation during steam injection. However, experimental data to confirm this feature are lacking.

## CONCLUSIONS

[49] A coupled thermo-hydro-mechanical modelling of SAGD showed how the different fronts (dilation, pressure, temperature and steam) develop and propagate with respect to time. The invasion of these fronts will impact 4D seismic monitoring because of the changes in seismic attributes (velocities, attenuations, etc.).

[50] The Ciz and Shapiro approach, a generalization of the Biot-Gassmann approach taking into account the viscoelastic behaviour of heavy oil, has been tested to model the oil sand velocity dispersion. The model has been calibrated on the velocities measured in laboratory and in situ (before injection), then has been used to calculate the *P* and *S*-wave velocities in the reservoir at seismic frequency (100 Hz). Until the vapour injection, temperature appears to be the dominant factor affecting the wave property response through its dominant effect on the viscosity and the properties of the heavy oil. The sudden appearance or disappearance of gas, if it occurs, is probably the strongest factor in changing the velocities.

# ACKNOWLEDGMENTS

The authors would like to thank M. P. Rasolofosaon and M. G. Renard for many useful recommendations and helpful comments.